\documentclass[fleqn,10pt]{wlscirep}
\usepackage[utf8]{inputenc}
\usepackage[T1]{fontenc}
\usepackage{graphicx}
\usepackage{amsmath}
\usepackage{longtable}
\usepackage{array}
\usepackage{float}
\usepackage{diagbox}
\usepackage{subcaption}
\usepackage{parskip}
\usepackage[numbers,super]{natbib}
\usepackage{multirow}
\usepackage{booktabs} 
\usepackage{diagbox}
\usepackage{hyperref}
\newcommand{\one}{\mathrm{I}}
\newcommand{\two}{\mathrm{I}\hspace{-1.2pt}\mathrm{I}}
\title{Predicting Artificial Neural Network Representations to Learn Recognition Model for Music Identification from Brain Recordings}

\author[1,*]{Taketo Akama}
\author[1]{Zhuohao Zhang}
\author[1,a]{Pengcheng Li}
\author[1,a]{Kotaro Hongo}
\author[1]{Hiroaki Kitano}
\author[1]{Shun Minamikawa}
\author[1]{Natalia Polouliakh}

\affil[1]{Sony Computer Science Laboratories, Inc, Tokyo, Japan}
\affil[a]{Work conducted when working as a research assistant}
\affil[*]{Corresponding author: Taketo Akama, taketo.akama@sony.com}

\begin{abstract}
Recent studies have demonstrated that the representations of artificial neural networks (ANNs) can exhibit notable similarities to cortical representations when subjected to identical auditory sensory inputs. In these studies, the ability to predict cortical representations is probed by regressing from ANN representations to cortical representations. Building upon this concept, our approach reverses the direction of prediction: we utilize ANN representations as a supervisory signal to train recognition models using noisy brain recordings obtained through non-invasive measurements.
Specifically, we focus on constructing a recognition model for music identification, where electroencephalography (EEG) brain recordings collected during music listening serve as input. By training an EEG recognition model to predict ANN representations—representations associated with music identification—we observed a substantial improvement in classification accuracy.
This study introduces a novel approach to developing recognition models for brain recordings in response to external auditory stimuli. It holds promise for advancing brain-computer interfaces (BCI), neural decoding techniques, and our understanding of music cognition. Furthermore, it provides new insights into the relationship between auditory brain activity and ANN representations.
\end{abstract}

\begin{document}

\flushbottom
\maketitle
\thispagestyle{empty}

\section*{Introduction}
Neural decoding in the audio domain refers to the technology for extracting or reconstructing audio stimuli or their features from brain recordings obtained when the audio stimulus is processed by the brain \cite{daly2023neural,pasley2012reconstructing,akbari2019towards,anumanchipalli2019speech,bellier2023music,hoefle2018identifying}. This technique provides insights into the neural localization and dynamics relevant to auditory stimuli and their features \cite{correia2014brain,bellier2023music,DiLiberto2022}. Furthermore, it enables the exploration of individual differences in cognition, such as those between experts and non-experts or native and non-native listeners \cite{DiLiberto2022,millet2022toward}. Neural decoding in the audio domain also has various research and industrial applications as part of brain-computer interface (BCI) systems, including speech production \cite{pasley2012reconstructing,akbari2019towards,anumanchipalli2019speech,defossez2023decoding}, music identification \cite{marion2021music,foster2018decoding,schaefer2011name,sternin2016classifying,lawhatre2020classifying,stober2015deep,sonawane2020guessthemusic,ramirez2022eeg2mel,ramirez2022imagebased,pandey2022music,hoefle2018identifying}, and music reconstruction \cite{DiLiberto2022,daly2023neural,bellier2023music,di_liberto2021accurate}.

Methods for measuring brain responses to auditory stimuli include invasive techniques such as electrocorticography (ECoG) and non-invasive approaches such as functional magnetic resonance imaging (fMRI), electroencephalography (EEG), magnetoencephalography (MEG), and near-infrared spectroscopy (NIRS) \cite{vogel2015assistive}. Among these, EEG is particularly suited for high-temporal-resolution time-series data due to its high sampling rate \cite{daly2019eeg,vogel2015assistive}. Its affordability, ease of use, and portability suggest significant potential for widespread adoption \cite{vogel2015assistive}. However, EEG faces persistent challenges, including high noise levels and signal attenuation from subcortical regions \cite{daly2019eeg,schirrmeister2017deep}.

Recently, studies have reported that artificial neural network (ANN) representations exhibit similarities to cortical representations in response to the same auditory sensory input, i.e., the same auditory stimulus \cite{millet2022toward,vaidya2022selfsupervised,tuckute2023many,oota2023neural}. These studies explore the feasibility of predicting cortical representations by regressing from ANN representations to cortical representations, thereby probing the alignment between these two types of representations.

\begin{figure}[H]
\centering
\includegraphics[width=0.7\textwidth]{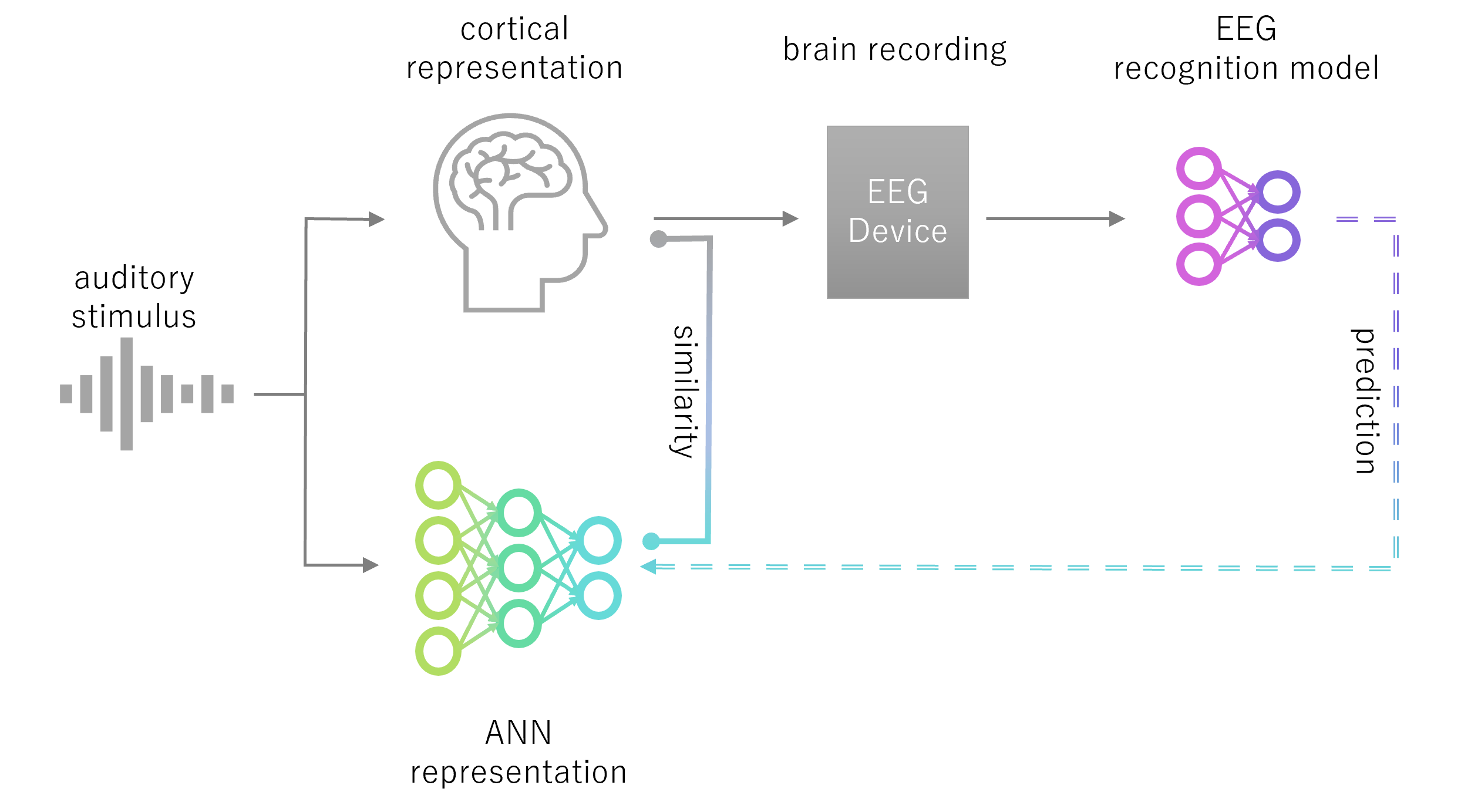}
\caption{Conceptual framework of our approach: predicting ANN representation to learn auditory EEG recognition model.
}
\label{fig:proposed framework}
\centering
\begin{minipage}{0.8\textwidth}
\footnotesize
When constructing a recognition model that uses brain recordings obtained in response to an auditory stimulus as input, the model is trained to predict the ANN representation obtained by inputting the same auditory stimulus into the ANN. This framework improves the performance of the recognition model by effectively utilizing the findings that cortical representations and ANN representations resemble each other when the same auditory stimulus is inputted.
\end{minipage}
\end{figure}

Our hypothesis posits that if auditory brain response recordings under ideal conditions resemble the representations of artificial neural networks (ANNs), then the information contained in ANN representations may be useful for complementing incomplete brain response recordings obtained under less-than-ideal conditions.
We attempt to validate this hypothesis by treating ANN representations as the target signals to be predicted, training a transformation function that complements the measured brain response information, and observing improvements in the accuracy of neural decoding models (recognition models) utilizing the complemented brain response information.
In prior studies evaluating the similarity between brain and ANN representations in response to auditory stimuli, cortical representations were predicted from ANN representations using small linear models, and a high prediction accuracy was interpreted as evidence of similarity. In contrast, our approach reverses this direction of prediction. Specifically, we train a transformation model to enable recorded auditory brain responses to predict ANN representations. Rather than merely evaluating similarity, our objective is to leverage the similarity for information complementation. For this reason, we employ larger, non-linear models.
The ultimate goal is to effectively train auditory neural decoding models (recognition models) even when noisy brain recordings are obtained through non-invasive measurements, by using ANN representations as the target signals to be predicted. Figure \ref{fig:proposed framework} illustrates the conceptual framework of our approach.

In this study, we focus on developing a recognition model for music identification using EEG brain recordings collected during music listening. By training a recognition model with EEG inputs to infer representations capable of predicting the ANN representations used for music identification, we observed substantial improvements in classification accuracy. Furthermore, the model demonstrated robust learning that was less dependent on the initialization of model weights.
The most significant improvement in classification accuracy was observed when assuming a brain response delay of approximately 200 ms to music stimuli. This aligns with recent scientific findings on brain response delays to music stimuli \cite{DiLiberto2022}, suggesting that the model effectively utilizes the similarity between ANN and brain representations.
Additionally, the model exhibited an ideal property where longer durations of EEG input resulted in higher music identification accuracy. We also report on the variations in accuracy based on individual differences and differences in the music itself, offering insights from a neuroscience perspective.

To predict ANN representations, various methods can be considered. In this work, we propose using a contrastive learning approach \cite{Oord2018RepresentationLW,pmlr-v182-zhang22a,pmlr-v139-radford21a}. 
Unlike conventional contrastive learning, our method incorporates two major techniques. The first technique involves simultaneously solving a classification task on both the brain representation branch and the ANN representation branch when using ANN representations as the supervisory signal. This dual-task setup reduces the learning of irrelevant features, generates supervisory signals specific to the target task, and focuses the model on learning features essential for the current task. The second technique is that the ANN representation branch is not trained to align with the brain representation branch. Aligning the ANN representation branch too closely with the brain branch risks diminishing the discriminative capability of the ANN representations for the target task due to the noise inherent in brain recordings. In practice, we confirmed through experiments that implementing this second technique improves music identification accuracy.

This study proposes a novel approach to recognition models for brain recordings in response to music auditory stimuli. It holds promise for advancing the understanding of cognitive mechanisms through neural decoding, contributing to the development of brain-computer interfaces (BCI), and offering valuable insights into the relationship between the human music perception and ANN processing of music.

\section*{Results}
This study utilizes the NMED-T dataset, comprising EEG data collected from 20 subjects as they listened to 10 unique songs \cite{Losorelli2017}. Each subject's EEG data was recorded while they listened to these songs, resulting in a dataset suitable for examining how neural responses align with specific musical features. In this experiment, we approached the task as a 10-class classification problem. The input data consists of EEG signals, while the output is the corresponding song ID. Given the 10 possible classes (one for each song), the chance level accuracy for this classification task is 0.1.

In this section, we present the findings from our model testing and evaluations. The analysis begins with optimizing the strength of predicting ANN representation to enhance model performance, followed by an assessment of robustness across various random seed values. We also adjust the model to consider delays between audio onset and participant perception. A comparison between 1D and 2D CNN models as well as learning techniques help identify the optimal configuration. After selecting the best model, we benchmark its performance against prior studies. Further evaluations are conducted on extended duration of EEG to assess flexibility. Finally, we examin model performance on individual songs and explore differences across participants, offering insights from both performance and neuroscientific perspectives.

\subsection*{Preliminary Model Testing}
We adopt the SampleCNN architecture as the base 1D CNN model\cite{lee2018samplecnn} both for audio and EEG models and extend it with two projectors: one for the classification task and another for the contrastive learning task.
In the pursuit of refining our model’s performance, parameter tuning was conducted focusing on optimizing the strength
of predicting ANN representation via the PredANN loss weight (see Methods section for more detail). We initiated this process by comparing weights of 0.01, 0.05, and 0.1 to clarify the optimal PredANN loss  weight, and the weight of 0.05 was identified as the optimal value for our model. To ensure robustness and precision in our findings we further scrutinized PredANN loss weights ranging from 0.03 to 0.07 at intervals of 0.01. The initial model was tested based on a refined optimal configuration comprising a PredANN loss weight of 0.05 and a seed of 42. The seed value determines the initial state of the model's parameters and different seeds can lead to different initialization of parameters impacting the reproducibility and the performance of the model. Seed 42 is a commonly utilized value in the scientific community due to its historical prevalence in computer science. Under this configuration, the model achieved a classification accuracy of 0.482. For comparative purposes, a baseline model with a PredANN loss weight of 0 was also evaluated, yielding a classification accuracy of 0.474. McNemar's test indicated a statistically significant difference between the models (p < 0.001), suggesting that the predicting ANN representation plays a critical role in the model's performance.

\subsection*{Robustness Testing}
To assess the robustness of our model against variations in random seed values, we conducted a series of experiments. While the primary evaluation was based on seed value 42, we broadened our investigation to include multiple seeds: 0, 1, and 2. This was executed for both our optimal PredANN loss weight of 0.05 and the baseline weight of 0. The objective was to discern whether our proposed model consistently outperforms the baseline across varied initializations. Results from the experiments were surprising for seed values 0 and 1. For these seeds, the baseline model's classification accuracy displayed a degree of unpredictability. In contrast, our proposed model exhibited more consistent and superior performance. Such outcomes suggest that our model beyond being adept at feature extraction might possess a capacity to handle intricate datasets and architectures. We conducted McNemar's test to compare the performance of our model with the baseline model. The results indicated significant differences for seed 0 (p<0.001), seed 1 (p<0.001), and seed 42 (p<0.001) whereas seed 2 (p=0.663) did not reveal a statistically significant difference. 

\begin{table}[ht]
\centering
\begin{tabular}{lcccc|cc}
\toprule
\diagbox{PredANN loss weight}{Seed} & 0 & 1 & 2  & 42 & Maximum & Average \\ 
\midrule
0.05 (our model) & \textbf{0.424} & \textbf{0.473} & \textbf{0.490} & \textbf{0.482} & \textbf{0.490} & \textbf{0.465} \\ 
0 & 0.100 & 0.100 & 0.486 & 0.474 & 0.486 & 0.324 \\ 
\bottomrule
\end{tabular}
\caption{Accuracy of proposed model for different seeds}
\label{tab:seed_accuracy}
\centering
\begin{minipage}{0.5\textwidth}
\footnotesize
With the PredANN loss weight of 0.05, the model achieved significantly higher accuracy for three out of four tested seeds: seed 0 (p<0.001), seed 1 (p<0.001), and seed 42 (p<0.001).
\end{minipage}
\end{table}

To provide a comprehensive evaluation, we considered two primary metrics: the maximum classification accuracy across all seeds and the average accuracy across the seed values. Both these metrics underscored the superior performance of our proposed model relative to the baseline.

\subsection*{Incorporating Time-delay}
We hypothesized that introducing relative delays between music auditory signal and its EEG encoding may enhance EEG classification accuracy. This was based on previous findings that suggested optimal EEG encoding of music occurs approximately 200ms post-stimulus onset \cite{Jagiello2019, DiLiberto2022}. To explore this, we introduced varied latencies between EEG and the music stimulus to evaluate model performance.

\begin{figure}[H]
\centering
\includegraphics[width=0.5\textwidth]{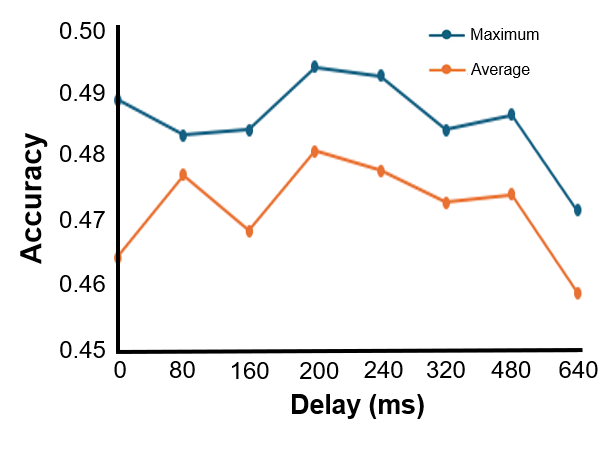}
\caption{The accuracy of different delays.}
\label{fig:evaluation_delay}
\centering
\begin{minipage}{0.5\textwidth}
\footnotesize
The line graph illustrates the variation in accuracy across different delay intervals, with the blue line representing the maximum method and the orange line denoting the average method. Both evaluation methods exhibited a peak at a delay of 200 ms. This aligns with prior research, suggesting that this delay corresponds to the typical human auditory reaction time of musical onset, which plays a crucial role in optimizing model performance.
\end{minipage}
\end{figure}

In our initial analysis, we examined delay intervals of 80 ms, 160 ms, 320 ms, and 640 ms, using seed values of 0, 1, and 2 to assess robustness. The model’s performance for these intervals is shown in the subsequent table, where we evaluated overall efficacy using both the \textbf{maximum} and \textbf{average} values across all seeds. This approach allowed us to capture both peak performance and consistent performance trends across different conditions. No clear peak performance emerged among these four delays. Thus, we further tested intermediate values, specifically 240 ms (between 160 ms and 320 ms) and 480 ms (between 320 ms and 640 ms). Results indicated that 240 ms yielded superior accuracy in both maximum and average metrics. To further refine the optimal delay, we examined 200 ms—the midpoint between 160 ms and 240 ms—and found it provided the highest accuracy, suggesting it as the optimal delay, which are consistent with previous studies. Table \ref{tab:delay_accuracy} presents classification accuracies across the delay intervals, with the 200 ms model demonstrating the best overall performance. Figure \ref{fig:evaluation_delay} further illustrates that, at 200 ms, the model achieved superior performance across all evaluation metrics, confirming its robustness.

We also performed McNemar's test to compare the results between delay 0 ms and delay 200 ms across all seeds. The test revealed significant differences for seed 0 (p<0.001) and seed 1 (p=0.0043), while no significant difference was found for seed 2 (p=0.306). 

\begin{table}[ht]
\centering
\begin{tabular}{lccccc}
\toprule
\diagbox{Delay}{Seed} & 0 & 1 & 2 & Maximum & Average \\ 
\midrule
0ms  & 0.424 & 0.473 & 0.490 & 0.490 & 
0.465 \\
80ms  & 0.467 & 0.483 & 0.484 & 0.484 & 0.478 \\ 
160ms & 0.459 & 0.464 & 0.485 & 0.485 & 0.469 \\ 
200ms & 0.483 & 0.467 & 0.495 & 0.495 & 0.482 \\ 
240ms & 0.478 & 0.464 & 0.494 & 0.494 & 0.479 \\ 
320ms & 0.460 & 0.477 & 0.485 & 0.485 & 0.474 \\ 
480ms & 0.488 & 0.464 & 0.473 & 0.488 & 0.475 \\ 
640ms & 0.473 & 0.446 & 0.460 & 0.473 & 0.459 \\ 
\bottomrule
\end{tabular}
\caption{Accuracy of time delay model for different seeds}
\label{tab:delay_accuracy}
\centering
\begin{minipage}{0.5\textwidth}
\footnotesize
This table presents the results for different delay settings, with the 200 ms delay yielding the highest accuracy. Statistical significance was observed in comparison to no delay  for seed 0 (p<0.001) and seed 1 (p=0.0043), whereas seed 2 did not exhibit statistical significance (p=0.306).
\end{minipage}
\end{table}

\subsection*{2D CNN vs 1D CNN}
After validating the 1D CNN model, we compared the 1D CNN with the 2D CNN. The 2D CNN architecture was selected based on prior research\cite{Ramirez-Aristizabal2022}. We tested 200ms delay with seed values 0, 1, and 2 and results showed that the 2D CNN outperformed  the 1D CNN. Additionally we compared the baseline models (2D CNN with PredANN loss weight 0) and obtained the expected results. As we mentioned in the Introduction, in our main proposed model, ANN representation model (music model) is not trained to align with the brain representation model (EEG model). Aligning
the ANN representation model (music model) closely with the brain representation model (EEG model) risks diminishing the discriminative capability of the ANN
representations for the target task due to the noise inherent in EEG recordings. Specifically, by applying the stop-gradient method, we ensure that the gradients of the PredANN loss, the purpose of which is to predict ANN representation from the EEG representation, are used to update the EEG model, but not the music model. To validate the effectiveness of our model, we also compare the results with and without the use of the stop-gradient method (stop-gradient-free model). 

To compare the performance between the 1D CNN and 2D CNN models, we conducted McNemar's test between the 1D CNN (PredANN loss weight 0.05) and the 2D CNN (PredANN loss weight 0.05). Additionally, we compared the 2D CNN (PredANN loss weight 0.05) with the baseline 2D CNN (PredANN loss weight 0) and the stop-gradient-free 2D CNN (PredANN loss weight 0.05).
The results showed that for all seeds (0, 1, and 2), the p-values for the comparison between 1D CNN and 2D CNN were less than 0.001, indicating significant differences, with the 2D CNN performing considerably better than the 1D CNN. When comparing the 2D CNN (PredANN loss weight 0.05) with the baseline 2D CNN (PredANN loss weight 0), significant differences were observed for seed 0 (p=0.035) and seed 2 (p=0.0013), but no significant difference was found for seed 1 (p=0.367). Lastly, in the comparison between the 2D CNN (PredANN loss weight 0.05) and the stop-gradient-free 2D CNN (PredANN loss weight 0.05), all seeds (0, 1, and 2) yielded p-values of less than 0.001.
\begin{table}[ht]
\centering
\begin{tabular}{lccccc}
\toprule
\diagbox{Model (delay 200ms)}{Seed} & 0 & 1 & 2 & Maximum & Average \\ 
\midrule
2D CNN, PredANN loss weight 0.05 (our best model)         & \textbf{0.662} & \textbf{0.622} & \textbf{0.588} & \textbf{0.662} & \textbf{0.624} \\ 
1D CNN, PredANN loss weight 0.05          & 0.487 & 0.465 & 0.494 & 0.494 & 0.482 \\ 
2D CNN, PredANN loss weight 0             & 0.537 & 0.589 & 0.516 & 0.589 & 0.547 \\ 
2D CNN, PredANN loss weight 0.05, stop-gradient-free & 0.648 & 0.529 & 0.315 & 0.648 & 0.497 \\ 
\bottomrule
\end{tabular}
\caption{Accuracy for different models (delay 200ms)}
\label{tab:cnn_comparison}
\centering
\begin{minipage}{0.8\textwidth}
\footnotesize
This table presents the results of various CNN comparisons, with the 2D CNN configured with a PredANN loss weight of 0.05 and applying stop-gradient to the music CNN achieving the best performance. This configuration was selected as the optimal model for further analysis.
\end{minipage}
\end{table}

\subsection*{Previous Study Comparison}
\begin{figure}[H]
\centering
\includegraphics[width=0.5\textwidth]{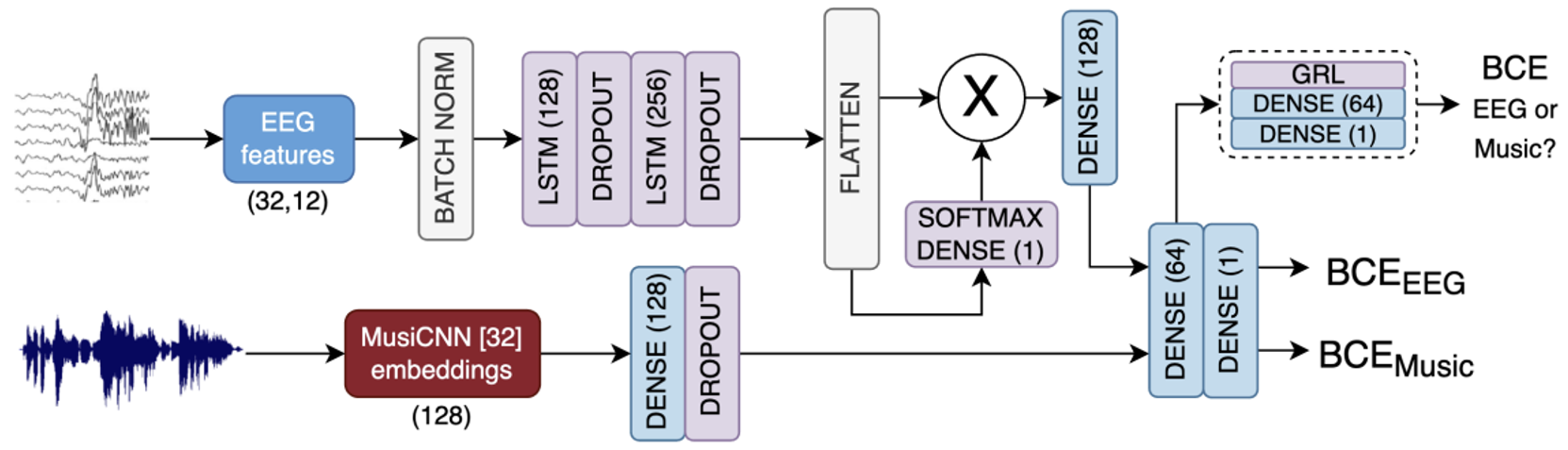}
\caption{Previous model structure}
\label{fig:previous_paper}
\centering
\begin{minipage}{0.5\textwidth}
\footnotesize
The network proposed in the previous study \cite{Avramidis2022} utilized a common layer to align EEG and music modalities, along with a GRL for domain adaptation.
\end{minipage}
\end{figure}

Subsequently, we compared the Gradient Reversal Layer (GRL)  method from previous studies\cite{Avramidis2022}. We retained our own 2D CNN model for feature extraction but modified the following layers and loss functions to be consistent with the prior research (Figure \ref{fig:previous_paper}). The aim was to compare our method with their GRL method. The results are shown in Table \ref{tab:previous_paper}. Although the method from previous studies achieved an accuracy above chance level (10\%), our proposed method demonstrated superior performance in the 10-class classification task. To evaluate model performance, we conducted McNemar's test, which revealed statistically significant differences for seed 0 (p < 0.001), seed 1 (p = 0.0016), and seed 2 (p < 0.001). These results suggest that our model performs significantly better on the current task.

\begin{table}[ht]
\centering
\begin{tabular}{lccc}
\toprule
\diagbox{Method}{Seed} & 0 & 1 & 2 \\ 
\midrule
Previous study & 0.178 & 0.144 & 0.155 \\ 
Our best model      & 0.662 & 0.622 & 0.588 \\ 
\bottomrule
\end{tabular}
\caption{Comparison with previous study}
\label{tab:previous_paper}
\centering
\begin{minipage}{0.5\textwidth}
\footnotesize
This table presents the comparison between the model in previous study and our proposed model. Our model outperformed the previous approach across all seeds, demonstrating statistically significant improvements: seed 0 (p<0.001), seed 1 (p=0.0016), and seed 2 (p<0.001).
\end{minipage}
\end{table}

\subsection*{Different Evaluation Length}
We subsequently explored classifying EEG with durations longer than 3 seconds. By applying overlapping 3-second sliding windows with a 1-second stride, we predicted each window separately to obtain results for the entire duration. Our best-performing model (2D CNN with PredANN loss weight 0.05, delay 200 ms, seed 0) was evaluated across duration of 3 to 7 seconds using three methods: mean (average prediction scores of all windows) max (highest score among windows) and majority (most frequent prediction among windows). Table \ref{tab:evaluation_length} presents these evaluation results indicating that the mean method achieved the best performance. 

We conducted McNemar's test to compare the 3-second and 7-second evaluation periods across all seeds (0, 1, and 2) and all methods (mean, max, and majority). The results showed that for both the max and mean methods, all seeds (0, 1, and 2) yielded p-values of less than 0.001, indicating significant differences between the evaluation lengths. However, for the majority method, only seed 1 demonstrated a significant difference (p = 0.0074), while no significant differences were observed for seed 0 (p = 0.819) and seed 2 (p = 0.746).

\begin{figure}[H]
\centering
\includegraphics[width=0.5\textwidth]{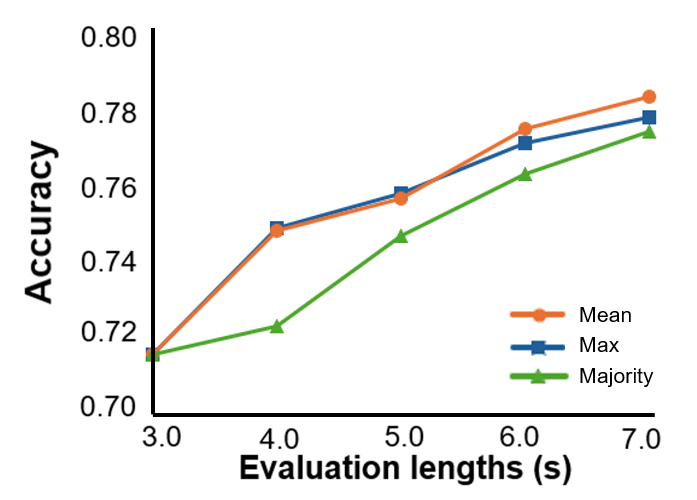}
\caption{The accuracy of longer evaluation length for three methods.}
\label{fig:evaluation_length}
\centering
\begin{minipage}{0.6\textwidth}
\footnotesize
The line graph illustrates the variation in accuracy across different evaluation lengths, with the orange line representing the mean method, the blue line denoting the maximum method, and the green line corresponding to the majority method. All three evaluation methods exhibit a consistent increase in accuracy as the evaluation length extends, highlighting the robustness and flexibility of the proposed model in handling longer evaluation sequences.
\end{minipage}
\end{figure}

Figure \ref{fig:evaluation_length} illustrates the results of the three methods. It can be observed that the accuracy of all methods increases with the evaluation duration, demonstrating that our model is capable of classifying durations longer than 3 seconds effectively. 
\begin{table}[ht]
\centering
\begin{tabular}{lccc}
\toprule
\diagbox{Length}{Method} & Mean & Max & Majority \\ 
\midrule
3s & 0.716 & 0.716 & 0.716 \\ 
4s & 0.748 & 0.749 & 0.723 \\ 
5s & 0.756 & 0.758 & 0.747 \\ 
6s & 0.774 & 0.771 & 0.763 \\ 
7s & 0.783 & 0.778 & 0.774 \\ 
\bottomrule
\end{tabular}
\caption{Accuracy for different evaluation lengths}
\label{tab:evaluation_length}
\centering
\begin{minipage}{0.5\textwidth}
\footnotesize
This table presents the results for different evaluation lengths, showing that accuracy improves as the evaluation length increases.
\end{minipage}
\end{table}

By using this approach, we enhance the model's flexibility. With training on a simple 3-second period, the model can effectively classify not only 3-second segments but also adapt to various input lengths without additional training. Our results also demonstrate the model’s generalizability, showing that it can maintain or even improve accuracy across inputs of different lengths. This approach enables sliding window processing, allowing the model to evaluate data in real time without requiring a full-duration signal, making it highly suitable for real-time EEG classification and neurofeedback systems.

\subsection*{Different Songs Evaluation}
\begin{figure}[H]
\centering
\includegraphics[width=0.43\textwidth]{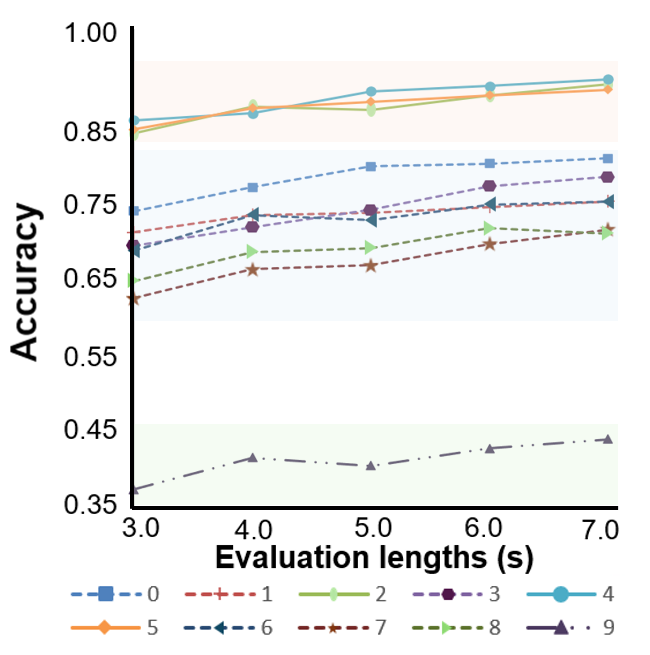}
\caption{Accuracies for different songs grouped into three}
\label{fig:song_evaluation}
\centering
\begin{minipage}{0.5\textwidth}
\footnotesize
The graph presents the results for individual songs, with distinct markers representing different songs. The songs are categorized into three groups: top, mid, and bottom, reflecting their respective contributions to model accuracy. This categorization suggests that distinctive musical elements may play a critical role in enhancing model performance.
\end{minipage}
\end{figure}

We used the best-performing method mean to evaluate different songs. Each song has unique musical features (such as rhythm and timbre) and we aimed to explore whether these features significantly impact classification accuracy. We hypothesized that unique musical characteristics could influence classification accuracy making some songs easier to classify. We evaluated durations from 3 to 7 seconds. Table \ref{tab:song_evaluation} and Figure \ref{fig:song_evaluation} show classification results for different songs at various lengths. Results indicate significant differences in classification accuracy. 

\begin{table}[ht]
\centering
\resizebox{\textwidth}{!}{
\begin{tabular}{lcccccccccc}
\toprule
\diagbox{Length}{Song\_id} & 0 & 1 & 2 & 3 & 4 & 5 & 6 & 7 & 8 & 9 \\ 
\midrule
3s & 0.754 & 0.725 & 0.859 & 0.707 & 0.877 & 0.864 & 0.700 & 0.636 & 0.659 & 0.377 \\ 
4s & 0.786 & 0.748 & 0.895 & 0.732 & 0.886 & 0.893 & 0.748 & 0.675 & 0.698 & 0.420 \\ 
5s & 0.814 & 0.752 & 0.891 & 0.755 & 0.916 & 0.902 & 0.741 & 0.680 & 0.704 & 0.409 \\ 
6s & 0.818 & 0.759 & 0.911 & 0.788 & 0.923 & 0.911 & 0.763 & 0.709 & 0.730 & 0.432 \\ 
7s & 0.825 & 0.766 & 0.925 & 0.800 & 0.932 & 0.918 & 0.766 & 0.729 & 0.723 & 0.445 \\ 
\bottomrule
\end{tabular}
}
\caption{Accuracies for different songs}
\label{tab:song_evaluation}
\centering
\begin{minipage}{0.5\textwidth}
\footnotesize
This table presents the results for different evaluation lengths across various songs, illustrating that accuracy varies depending on the song. 
\end{minipage}
\end{table}

All songs were categorized into three distinct groups based on classification accuracy. Songs no. 2, no. 4, and no. 5 demonstrated consistently high accuracy rates, all exceeding 85\%. These songs, which are highlighted in the orange group (the solid lines), likely contain unique or easily distinguishable musical elements that facilitate more precise classification. On the other end of the spectrum, Song no. 9 exhibited significantly lower accuracy compared to other songs and is therefore highlighted in the green group (a dot-dash line). This lower performance may indicate the presence of less distinct auditory features, posing a challenge for accurate classification. The remaining songs, grouped as the middle range and highlighted in blue (the dashed lines), achieved moderate accuracy levels. These songs may possess characteristics that are somewhat distinguishable but not as pronounced as those in the high-accuracy group. This distribution aligns with our hypothesis that songs with unique and prominent musical elements tend to yield higher classification accuracy. The analysis provides insight into how the distinctiveness of musical features can influence accuracy, which is explored in the discussion section.

\subsection*{Individual Evaluation}
\begin{figure}[H]
\centering
\hspace{8pt}
\includegraphics[width=0.45\textwidth, height=0.45\textwidth]{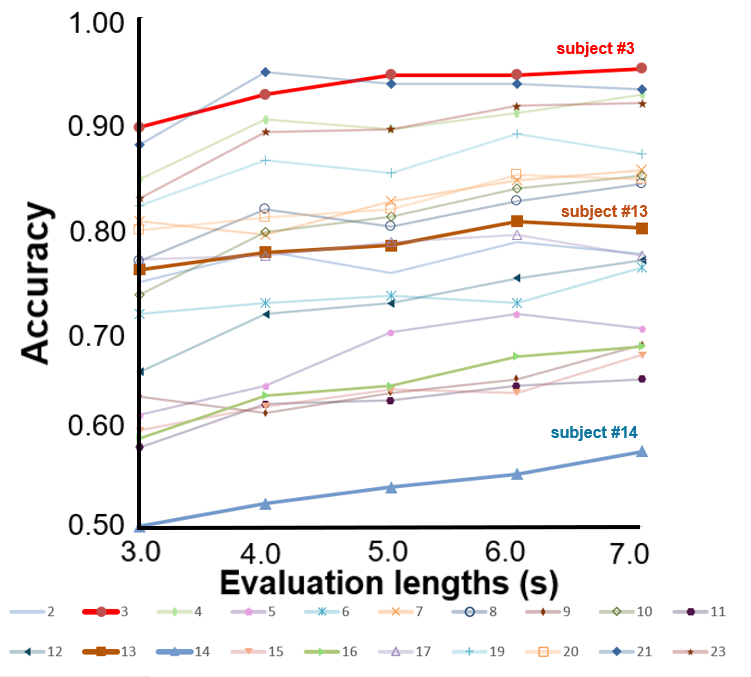}
\caption{Accuracies for different individuals in the graph}
\label{fig:individual_evaluation}
\centering
\begin{minipage}{0.5\textwidth}
\footnotesize
The graph illustrates the results for individual subjects, with distinct markers representing each subject. Three representative subjects were selected for further analysis: the best-performing subject \#3, an average-performing subject \#13, and the lowest-performing subject \#14. These results highlight the impact of individual differences on model performance.
\end{minipage}
\end{figure}

We then investigated individual performance hypothesizing that personal differences would be evident in the results. For instance, individuals skilled in music might achieve higher accuracy across more songs while some may struggle to identify features even in easily classifiable songs. Using the mean method, we evaluated durations from 3 to 7 seconds. Results shown in Table \ref{tab:individual_evaluation} and Figure \ref{fig:individual_evaluation} support our hypothesis highlighting significant individual differences in classification accuracy.

\begin{table}[ht]
\centering
\resizebox{\textwidth}{!}{
\begin{tabular}{lcccccccccccccccccccc}
\toprule
\diagbox{Length}{Subject} & 2 & 3 & 4 & 5 & 6 & 7 & 8 & 9 & 10 & 11 & 12 & 13 & 14 & 15 & 16 & 17 & 19 & 20 & 21 & 23 \\ 
\midrule
3s & 0.742 & 0.896 & 0.845 & 0.611 & 0.711 & 0.803 & 0.763 & 0.629 & 0.730 & 0.579 & 0.654 & 0.755 & 0.500 & 0.595 & 0.588 & 0.765 & 0.818 & 0.794 & 0.879 & 0.826 \\ 
4s & 0.773 & 0.929 & 0.904 & 0.639 & 0.721 & 0.789 & 0.814 & 0.612 & 0.792 & 0.621 & 0.711 & 0.772 & 0.523 & 0.619 & 0.630 & 0.769 & 0.864 & 0.807 & 0.951 & 0.891 \\ 
5s & 0.752 & 0.948 & 0.894 & 0.693 & 0.729 & 0.823 & 0.797 & 0.633 & 0.808 & 0.625 & 0.721 & 0.779 & 0.539 & 0.636 & 0.640 & 0.782 & 0.851 & 0.815 & 0.940 & 0.894 \\ 
6s & 0.783 & 0.948 & 0.910 & 0.711 & 0.721 & 0.844 & 0.823 & 0.646 & 0.835 & 0.639 & 0.746 & 0.803 & 0.552 & 0.633 & 0.669 & 0.789 & 0.890 & 0.849 & 0.940 & 0.917 \\ 
7s & 0.770 & 0.955 & 0.929 & 0.696 & 0.757 & 0.854 & 0.840 & 0.680 & 0.848 & 0.646 & 0.764 & 0.796 & 0.575 & 0.670 & 0.679 & 0.769 & 0.870 & 0.845 & 0.934 & 0.920 \\ 
\bottomrule
\end{tabular}
}
\caption{Accuracies for different individuals}
\label{tab:individual_evaluation}
\centering
\begin{minipage}{0.5\textwidth}
\footnotesize
This table presents the results for different evaluation lengths across individuals, illustrating that individual differences significantly impact model performance. 
\end{minipage}
\end{table}

\begin{figure}[H]
\centering
\begin{subfigure}{0.33\textwidth}
    \centering
    \includegraphics[width=\linewidth]{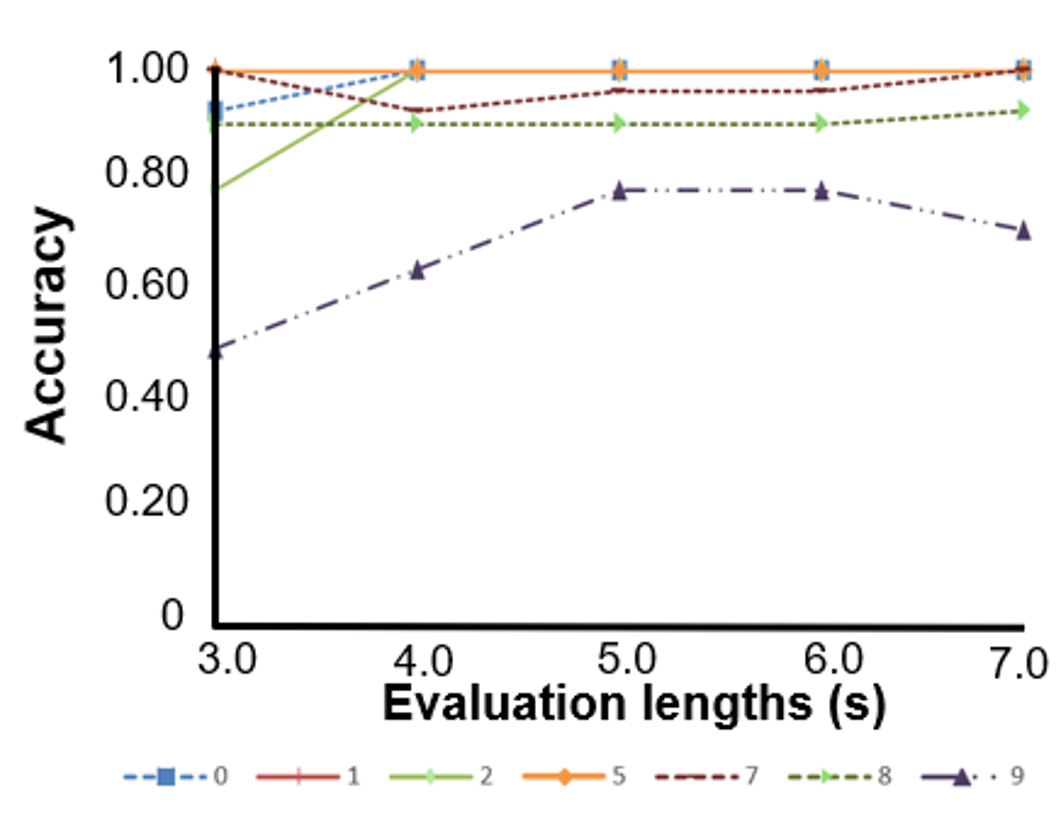}
    \caption{Subject no.3}
    \label{fig:subfig1}
\end{subfigure}
\begin{subfigure}{0.33\textwidth}
    \centering
    \includegraphics[width=\linewidth]{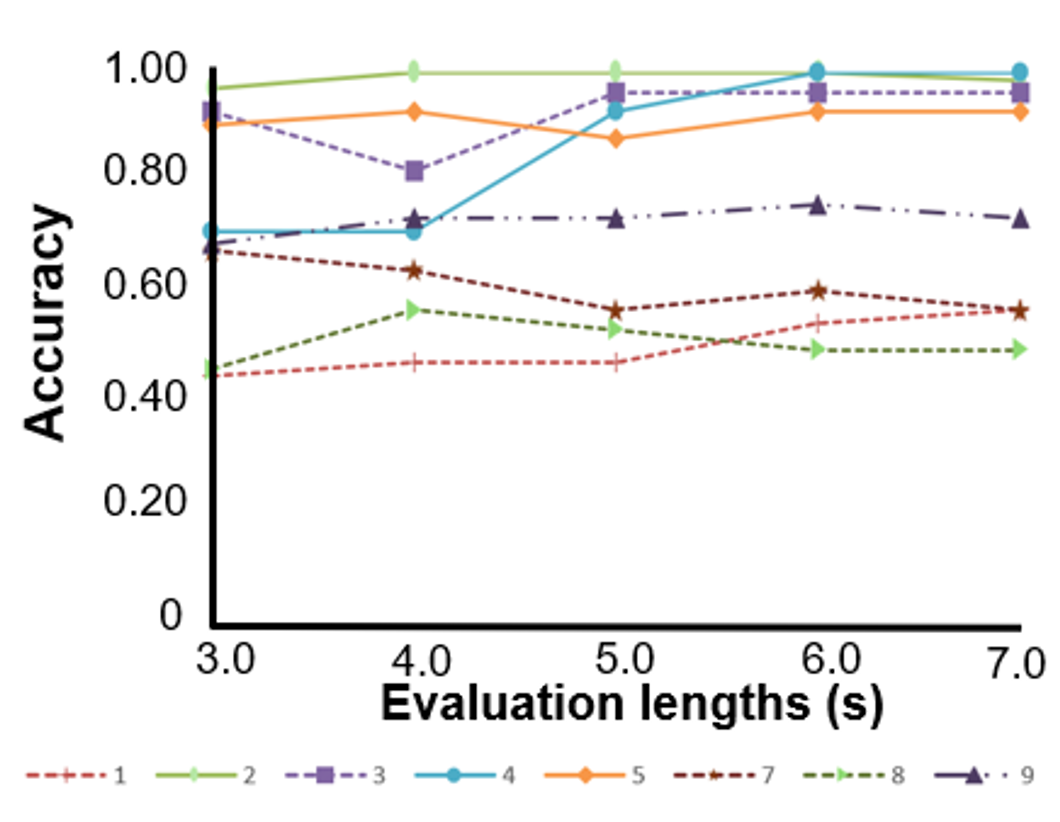}
    \caption{Subject no.13}
    \label{fig:subfig2}
\end{subfigure}
\begin{subfigure}{0.33\textwidth}
    \centering
    \includegraphics[width=\linewidth]{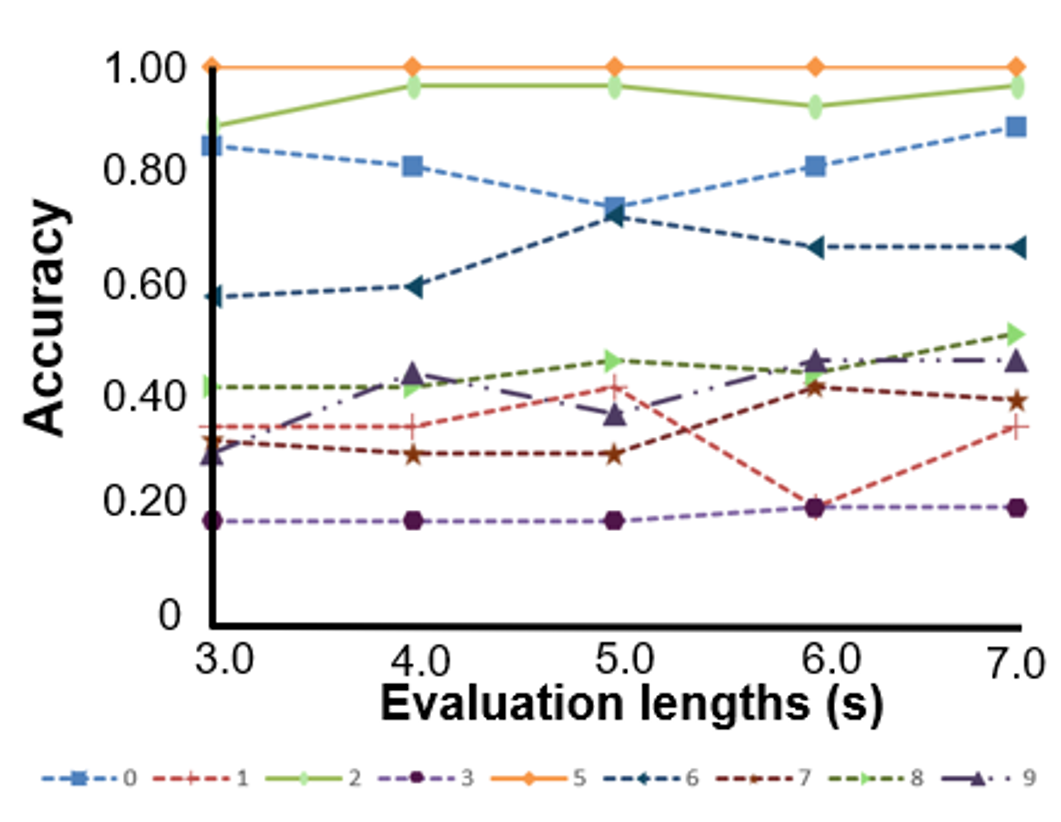}
    \caption{Subject no.14}
    \label{fig:subfig3}
\end{subfigure}
\caption{Individual results for (a) top group, (b) middle group, (c) bottom group}
\label{fig:combined}
\centering
\begin{minipage}{0.8\textwidth}
\footnotesize
The graph provides an individual analysis for the three selected subjects, with distinct markers representing different songs. The markers are consistent with those used in Figure \ref{fig:song_evaluation}. 
(a) Subject \#3: High accuracy was achieved across nearly all songs, indicating robust performance. (b) Subject \#13: Songs in the top group yielded higher accuracy, whereas other songs demonstrated comparatively lower accuracy levels. (c) Subject \#14: A more diverse range of accuracies was observed, reflecting greater variability in performance across songs.
This analysis underscores the varying impact of musical elements and individual differences on model accuracy. 
\end{minipage}
\end{figure}

To further analyze individual performance across different songs, we categorized subjects into three groups based on accuracy: top, middle, and bottom. We selected one subject from each group for detailed analysis: subject 3 (top), subject 13 (middle), and subject 14 (bottom). As shown in Figure \ref{fig:combined}, the high-accuracy subject demonstrated consistent performance across songs with most accuracies above 80\%. The middle group subject showed greater variability ranging from 40\% to 100\%. The bottom group subject displayed the most variability with accuracies as low as 20\%. Interestingly, songs 2, 4, and 5 had high accuracy across all groups suggesting that certain song features may outweigh individual abilities (detailed analysis is provided in the Discussion section).

\section*{Discussion}
\subsection*{Modeling the Relationship Between Auditory Stimuli and EEG}

Various studies have explored methods to model the relationship between audio stimuli or their features and brain recordings through mathematical functions. These approaches can be broadly categorized into regression, generation, and embedding into a common space. The objectives of these studies vary, including understanding brain function, retrieval, decoding, and elucidating the relationship between ANN representation and brain representation.

Let us first discuss regression. To capture complex relationships  rather than observing ERP (Event-Related Potential), linear models have been used to examine the relationship between auditory stimuli and brain recordings. Modeling from stimuli to recordings is referred to as a forward model (temporal response functions; TRF), while the reverse direction from recordings to stimuli is referred to as a backward model (backward temporal response functions; bTRF) \cite{crosse2016mtrf}. Forward models have been used to investigate what aspects of stimuli are encoded in EEG. For instance, regression from auditory stimuli or their features to brain recordings has been employed to identify which audio features are important for brain responses \cite{DiLiberto2022,bellier2023music}. Similarly, studies have shown that ANN audio representations resemble brain representations by linearly regressing from ANN representation to brain recordings and analyzing their correlations \cite{millet2022toward,vaidya2022selfsupervised,tuckute2023many,oota2023neural}.

Next, regarding generation, some studies aim to generate realistic music audio directly from brain recordings \cite{postolache2024naturalistic,denk2023brain2music,daly2023neural}. Unlike regression, these approaches often assume a one-to-many mapping for music generation and focus on outputting the audio signal itself rather than its features, prioritizing realism over interpretability.

Finally, methods categorized under embedding into a common space aim to learn a shared space between stimuli and brain recordings for tasks such as retrieval or stimuli reconstruction \cite{yu2018deep,defossez2023decoding}. While our study is closer to this category, it also simultaneously addresses a target classification task. Our objectives include decoding, understanding brain function, and elucidating the relationship between ANN representation and brain representation.

The above-mentioned prior studies, however, were not designed for downstream tasks (e.g., classification) based on representation learning for brain recordings, as is the focus of our study. Although not learning a one-to-one relationship between auditory stimuli and EEG as our study does, the closest concept is found in a study that uses domain adaptation for emotion classification, where the distributional information of music audio representations is reflected in that of EEG representations \cite{avramidis2022enhancing}. In our experiments, we demonstrated the superiority of our approach over this method. By leveraging the one-to-one relationship, our method provides a significantly richer supervisory signal, which we believe underpins the advantage of our proposed approach.

\subsection*{Another Interpretation of Our Framework}

As stated in the Introduction, the framework proposed in this study leverages the similarity between auditory cortical representations and ANN representations in a reverse manner. Specifically, it aims to effectively extract essential information from recordings of auditory cortical representations, even when noise is present.

Here, we offer an alternative interpretation of what our framework accomplishes. Auditory brain responses can be understood as signals transformed from auditory stimuli into neural activity. Indeed, methods such as temporal response functions (TRF) aim to approximate this transformation through a mathematical function relationship, thereby facilitating an understanding of the connection between auditory stimuli and their brain responses \cite{crosse2016mtrf}. From this perspective, our proposed method can be interpreted as performing an inverse transformation, akin to backward temporal response functions (bTRF), where brain responses are transformed back to stimuli. This process attempts to recover stimuli-related information and assists in making inferences about the stimuli.

However, instead of performing an inverse transformation into the data space of the auditory stimuli, our framework transforms the brain responses into the feature space of ANNs. This allows us to obtain high-level feature-based supervisory signals related to the stimuli while avoiding the challenging problem of detailed low-level reconstructions of stimuli that are not directly relevant to the task at hand.

\subsection*{Previous Paper Comparison}
The key distinction between our approach and that of Avramidis et al. \cite{Avramidis2022} lies in the alignment method. While their model employs a `set-to-set' alignment by integrating music and EEG features into a common layer and distribution matching, our model utilizes a `point-to-point' alignment through predicting ANN representation using contrastive learning. Our method aligns individual data points directly rather than aligning the two set of points across modalities, which we hypothesize is the main driver of our model's superior performance. Our `point-to-point' approach allows for more precise feature mapping, leading to enhanced classification accuracy in multi-class tasks like ours.
Moreover, our prediction method with dedicated projection heads effectively captures the nuanced relationships between EEG and music data, avoiding the distortion that can occur when forcing two different modalities to align in the classification head, as seen in the prior method.

In their study, Avramidis et al. \cite{Avramidis2022} set the domain loss weight to 0.1 after determining the optimal weight for their model. Notably, we adjusted the domain loss weight to 0, resulting in a significant performance improvement. Further experimentation with weights of 0.01, 0.05, and 0.001 showed that the best performance was achieved at 0.01, while both 0.05 and 0.001 performed similarly to a weight of 0, all exceeding the performance of the 0.1 setting. These results indicate that the domain loss is not effective in our multi-class Song ID classification task.

From a neuroscientific perspective, the differences in datasets used in our study and theirs are likely a critical factor in the observed performance discrepancies. Their research utilized the DEAP dataset, which is focused on emotion analysis \cite{koelstra2011deap}, engaging the frontal cortex, which is primarily responsible for processing emotional stimuli. In contrast, our study used the NMED-T dataset, focusing on musical rhythms processed mainly in the temporal cortex, linked to auditory and rhythmic perception.

\subsection*{Longer EEG Evaluation}
Our analysis indicated that the mean and max methods yielded better classification results compared to the majority method. The max method, by taking the highest prediction value among all windows, minimizes the impact of less distinct windows on the final result. Similarly, the mean method averages prediction values across all windows, enabling the influence of windows with higher accuracy to have a more significant impact.
Conversely, the majority method treats each window equally by selecting the most frequently occurring prediction, which may reduce the influence of windows with higher individual accuracy.

\begin{table}[ht]
\centering
\resizebox{\textwidth}{!}{
    \begin{tabular}{lc}  
        \toprule
        \textbf{Window} & \multicolumn{1}{c}{\textbf{Predicted Values}} \\
        \midrule
        Window 1 & [1.3501e-01, 5.5643e-07, 1.7565e-02, 2.1875e-03, 1.0757e-02, 4.0120e-03, 4.6196e-04, 9.5319e-03, \textbf{8.1936e-01}, 1.1074e-03] \\
        Window 2 & [\textbf{3.7487e-01}, 1.0177e-05, 8.1639e-03, 2.0318e-02, 1.2092e-01, 1.7365e-04, 3.3524e-03, 1.2068e-01, 3.3947e-01, 1.2045e-02] \\
        Window 3 & [7.0974e-03, 3.1501e-09, 4.2614e-05, 2.0529e-05, 1.6606e-04, 3.4932e-07, 1.3441e-06, \textbf{9.9258e-01}, 9.3455e-05, 2.0433e-06] \\
        \bottomrule
    \end{tabular}
}
\caption{Predicted values after softmax for each window}
\label{tab:softmax_predictions}
\centering
\begin{minipage}{0.5\textwidth}
\footnotesize
This table provides an example of a 5-second evaluation, presenting all three evaluation windows within the 5-second interval. The bold numbers represent the highest prediction value within each window.
\end{minipage}
\end{table}

\begin{table}[ht]
\centering
\begin{tabular}{lc}  % 
    \toprule
    \textbf{Step} & \textbf{Predicted Class} \\
    \midrule
    Window 1 & 8 \\
    Window 2 & 0 \\
    Window 3 & 7 \\
    \cmidrule(lr){1-2}  % 
    \textbf{Final Prediction} & 8 \\
    \bottomrule
\end{tabular}
\caption{Majority method}
\label{tab:majority_voting}
\centering
\begin{minipage}{0.5\textwidth}
\footnotesize
The table presents the predicted class for each evaluation window, along with the final prediction aggregated from all windows as the overall prediction for the 5-second period using the majority method.
\end{minipage}
\end{table}

\begin{table}[ht]
\centering
\resizebox{\textwidth}{!}{
    \begin{tabular}{c c}  % 只有两列
        \toprule
        \multicolumn{1}{c}{\textbf{Step}} & \multicolumn{1}{c}{\textbf{Predicted Values}} \\
        \midrule
        Max(Window 1, Window 2, Window 3) & [3.7487e-01, 1.0177e-05, 1.7565e-02, 2.0318e-02, 1.2092e-01, 1.7365e-04, 3.3524e-03, \textbf{9.9258e-01}, 8.1936e-01, 1.2045e-02] \\
        \cmidrule(lr){1-2}  
        \textbf{Final Prediction} & \multicolumn{1}{c}{7} \\
        \bottomrule
    \end{tabular}
}
\caption{Max method}
\label{tab:max_value}
\centering
\begin{minipage}{0.5\textwidth}
\footnotesize
The table presents the final predicted values aggregated from each evaluation window using the maximum method, along with the final predicted song number derived from these values for the 5-second evaluation period.
\end{minipage}
\end{table}

As an example, we output a 5-second evaluation period with a true label of 7 divided into three windows. Each window outputs prediction scores for 10 songs (Table \ref{tab:softmax_predictions}). The highest score for each window is highlighted in bold. Table \ref{tab:majority_voting} shows the majority method. We first determine the final prediction for each window which are 8, 0, and 7 respectively. We then select the most frequent prediction. In this case, the counts are equal so we choose the first value 8 as the final 5-second prediction result. The prediction result is incorrect. However for the max method, as shown in Table \ref{tab:max_value}, we first identify the highest score in each window to obtain the prediction results for all songs. Then we select the highest score among the 10 songs (highlighted in bold). In this example, the score for label 7 in window 3 is significantly high. Using the max method, the influence of window 3 is maximized resulting in a final prediction of 7 which is correct.

Our results demonstrated that accuracy increased with the length of the evaluation period. We speculate that the different classification accuracies over consecutive windows is due to the difference in informative features, such as unique melodies or rhythms as well as confusing features or EEG noise. 
This approach offers flexibility, enabling the model to classify not only the 3-second segments used in training but also to adapt seamlessly to longer input lengths without requiring additional training. Our results show that a model trained on 3-second segments generalizes well across various input lengths, often maintaining or even improving classification accuracy. 

The sliding window method facilitates real-time EEG processing by allowing the model to make continuous evaluations without needing the full signal duration upfront. This capability is particularly valuable for applications requiring immediate feedback, such as brain-computer interfaces, adaptive neurofeedback systems, or real-time auditory processing.

\subsection*{Different Songs and subjects}
We found that songs 2, 4, and 5 consistently achieved high classification accuracy, regardless of the participant's level.  
We analyzed the characteristics of each song and found that songs with higher accuracy tended to feature electronic elements and unique melodies. We hypothesize that this increased accuracy is due to the presence of synthetic electronic sounds and distinctive rhythmic patterns, which evoke more pronounced brain responses to these unusual elements. Recent research supports this hypothesis by demonstrating that familiarity with music can lead to varying levels of brain responses. Studies indicate that distinctive elements, such as unique or unfamiliar musical features, tend to evoke more pronounced brainwave activity.\cite{Li2024,Ding2024}

For instance, song 4 \textit{Lebanese Blonde}, which blends Middle Eastern instruments with electronic effects, creates an exotic and non-traditional melodic scale. This combination likely produces complex neural responses, as the brain processes the unfamiliar and intricate auditory patterns. These distinct neural signatures might make it easier for the classification model to differentiate this song from others, contributing to its higher accuracy.

We also examined individual performance and found that certain participants consistently achieved higher accuracy compared to others. This variability may be attributed to individual differences in music perception, where distinct levels of brain activity are elicited based on factors such as musical training, attentional focus, or psychological states \cite{Ding2024}. These individual differences likely contribute to the varying neural responses, impacting the accuracy of music-related tasks across participants. \cite{MartinezMolina2016}
% (Chiyo) Rewrite from Related work
The connection between music and brain activity has been widely acknowledged. Previous studies have demonstrated that musical tempo and rhythm can evoke distinct neural oscillations \cite{hoefle2018identifying, Stober2016, Rivera-Tello2023}, while attention to specific musical instruments can lead to unique EEG patterns \cite{Cantisani2019}. Building on these findings, our results suggest that variations in rhythmic and melodic structures, as well as instrumentation, influence EEG responses and subsequent classification performance. Notably, song 4 \textit{Lebanese Blonde} achieved the highest classification accuracy, which may be attributed to its particularly distinctive melody, complex rhythms, and unique instrumentation, all of which could elicit more salient and recognizable EEG features. 

Together, these findings indicate that both the intrinsic properties of songs and individual perceptual factors play key roles in EEG-based music classification accuracy.

\section*{Methods} 
\subsection*{Dataset and Preprocessing}
\subsubsection*{Dataset Information}
The dataset used in this study is Naturalistic Music EEG Dataset—Tempo (NMED-T), which is an open dataset collected from 20 participants, who listened to a selection of 10 commercially available songs \cite{Losorelli2017}. The dataset contains EEG data collected from 128 scalp electrodes during natural music appreciation. The EEG data in this study was originally sampled at 1000Hz. To optimize computational efficiency without losing significant information and focus on the most relevant frequency bands, it was subsequently down-sampled to 125Hz for our study.

\subsubsection*{Dataset Splitting Method}
The NMED-T dataset is composed of recordings from 20 subjects, each of whom listened to 10 full-length songs. This leads to a total of 200 files within the dataset. To maintain uniformity across all data, we standardized the length utilizing the maximal shared duration of four minutes for our analysis. We utilized the initial 4-minute segments from the recordings and further partitioned these segments into 30-second excerpts. To ensure that the training and validation sets preserved the original distribution of songs, we applied stratified sampling during dataset splitting. Specifically, we used the \textit{train\_test\_split} function from the \textit{sklearn.model\_selection} module\cite{scikit-learn}, with the \textit{stratify} parameter set to the song labels. For further details, see the provided code file \textit{preprocessing\_eegmusic\_dataset.py}. This approach guarantees that each song's proportionate representation is maintained across both the training and validation sets. For example, if a particular song accounts for 10\% of the entire dataset, it would also represent approximately 10\% of both the training and validation sets. This stratified method ensures that the diversity and proportions inherent to the complete dataset are effectively mirrored in both subsets, thereby enhancing the reliability and validity of our experimental outcomes. The dataset was divided into training and validation sets with a 75:25 ratio, ensuring consistent representation of each song across both sets. 

\subsubsection*{Preprocessing}
In this study, following Défossez et al. \cite{Defossez2022}, we employed the \textit{RobustScaler} for normalizing EEG data, followed by a \textit{clamp} operation to ensure data stability. Specifically, the \textit{RobustScaler} uses the median and interquartile range for scaling, which minimizes the influence of outliers on the data distribution. This approach is particularly suitable for EEG data, as EEG signals may contain sporadic noise and artifacts. For further details, see the \textit{normalize\_EEG\_4 } function in the provided code file \textit{preprocessing\_eegmusic\_dataset.py}. We use two steps to normalize the data. First, the \textit{RobustScaler} is applied individually to each channel, producing standardized data. Subsequently, a \textit{clamp} operation is used to restrict the values within a specified range (±20). If a value is lower than this range it's set to the minimum limit. If a value is higher than the range, it's set to the maximum limit. 

\subsection*{Model}
\subsubsection*{Model Architecture and Losses }\label{sec:proposed_model}
\begin{figure}[H]
\centering
\includegraphics[width=0.7\textwidth]{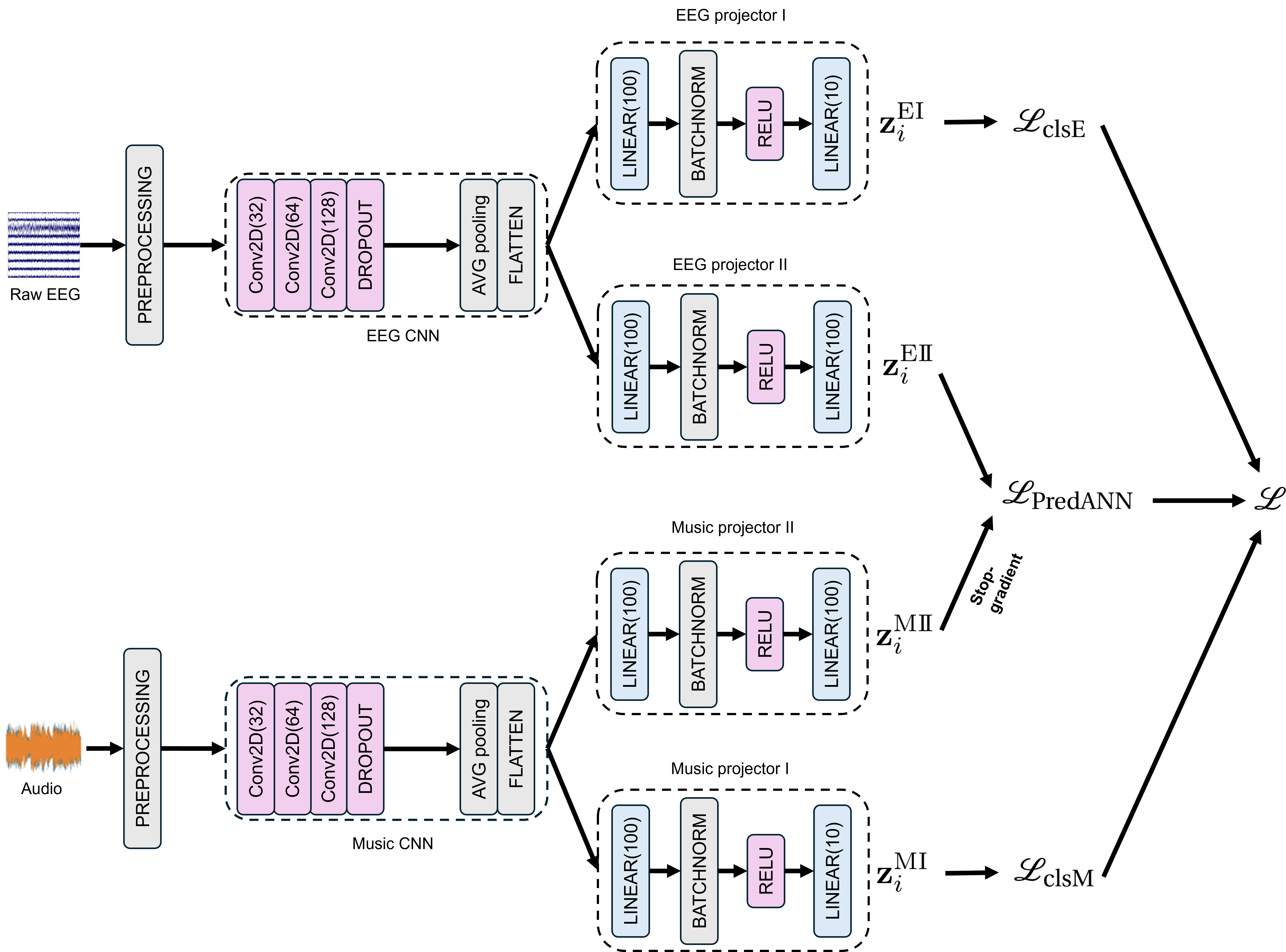}
\caption{The proposed model.}
\label{fig:proposed model}
\centering
\begin{minipage}{0.8\textwidth}
\footnotesize
The graph illustrates the structure of the proposed model. Two separate 2D CNNs are employed to process music and EEG data independently. The outputs include individual losses for music and EEG, along with a contrastive loss for learning the relationship between the two modalities.
\end{minipage}
\end{figure}

The proposed model consists of two distinct but structurally identical CNN-based encoders: one for processing raw EEG data and another for processing audio data. Each encoder leverages 2D CNNs to extract modality-specific features. 
To maintain consistency in processing both EEG and music data, we applied \textit{padding=1} in all convolutional layers. For EEG data, this ensures that spatial and temporal features are extracted uniformly without reducing the size of spatial and temporal lengths. For music data, \textit{padding=1} allows one-dimensional sequences to be seamlessly processed by 2D convolutions without altering its original length. This unified padding strategy simplifies model design and supports effective feature extraction across modalities (Figure \ref{fig:proposed model}).

The outputs from each encoder are then directed into two separate projectors: Projector I, focused on classification of 10 songs, and Projector II, dedicated to contrastive learning. The projectors outputs different type of embeddings for both EEG and music data: song embeddings from Projector I (for classification tasks) and feature embeddings from Projector II (for contrastive learning).
The model incorporates both Cross-Entropy (CE) loss for classification and PredANN loss for auxiliary representation learning. The CE loss is computed on the song embeddings to optimize the classification predictions for each class label. Meanwhile, the PredANN loss calculates similarity between feature embeddings from both EEG and music encoders, encouraging shared feature space alignment between the two modalities. The stop-gradient operation is applied to the music feature embeddings, preventing gradients from backpropagating through the music encoder in the contrastive task. The final loss function is a weighted combination of three components: the EEG classification loss, the music classification loss, and the PredANN loss. This cumulative loss function drives joint optimization, allowing the model to simultaneously learn discriminative features for classification and representations for contrastive learning. Formally, let $\{\mathbf{z}_i^{\text{E}\one}\}_{i=0}^{B}$, $\{\mathbf{z}_i^{\text{E}\two}\}_{i=0}^{B}$, $\{\mathbf{z}_i^{\text{M}\one}\}_{i=0}^{B}$, and $\{\mathbf{z}_i^{\text{M}\two}\}_{i=0}^{B}$ be outputs of EEG projector I, EEG projector I\hspace{-1.2pt}I, Music projector I, and Music projector I\hspace{-1.2pt}I, respectively, in a mini-batch of size $B$ when training the model. The classification loss for EEG and music are defined as 
\begin{equation}
\mathcal{L}_\mathrm{clsE} := \sum_{i=0}^{B}\texttt{CE}\left(\mathbf{z}_i^{\text{E}\one}, c_{i}^{\text{E}}\right)\;\; \text{and}\;\; \mathcal{L}_\mathrm{clsM} := \sum_{i=0}^{B}\texttt{CE}\left(\mathbf{z}_i^{\text{M}\one}, c_{i}^{\text{M}}\right),
\end{equation}
respectively, where $\texttt{CE}$ denotes cross entropy loss, and $c_{i}^{\text{E}}$ and $c_{i}^{\text{M}}$ denote classification labels of EEG and music, respectively. The PredANN loss is defined as
\begin{equation}
\mathcal{L}_\mathrm{PredANN} := - \sum_{i=0}^{B}\left(\log \frac{\exp{\left(\texttt{sim}\left(\texttt{sg}(\mathbf{z}_i^{\text{M}\two}),\mathbf{z}_i^{\text{E}\two}\right)/\tau\right)}}{\sum_{j=0}^{B}\exp{\left(\texttt{sim}\left(\texttt{sg}(\mathbf{z}_i^{\text{M}\two}),\mathbf{z}_j^{\text{E}\two}\right)/\tau\right)}}
+
\log \frac{\exp{\left(\texttt{sim}\left(\texttt{sg}(\mathbf{z}_i^{\text{M}\two}),\mathbf{z}_i^{\text{E}\two}\right)/\tau\right)}}{\sum_{j=0}^{B}\exp{\left(\texttt{sim}\left(\texttt{sg}(\mathbf{z}_j^{\text{M}\two}),\mathbf{z}_i^{\text{E}\two}\right)/\tau\right)}}
\right),
\end{equation}
where $\texttt{sim}(\cdot,\cdot)$, $\texttt{sg}(\cdot)$, and $\tau$ denote cosine similarity, stop-gradient operation, and the temperature parameter, respectively. The PredANN loss is based on the maximization of mutual information between EEG and music embeddings, rather than the maximization of likelihood of conditional probabilities. In practice, we use InfoNCE loss that approximates the negative mutual information, proposed in the context of Contrastive Predictive Coding (CPC) of speech or image \cite{Oord2018RepresentationLW}. The use of InfoNCE loss between different modalities has been first explored in ConVIRT in image and text domain \cite{pmlr-v182-zhang22a} and later popularized by CLIP \cite{pmlr-v139-radford21a}. Unlike these prior works, our PredANN loss introduces stop-gradient operations $\texttt{sg}(\cdot)$ and applied to EEG and music domain. The final loss function is
\begin{equation}
\mathcal{L} := \mathcal{L}_\mathrm{clsE} + \mathcal{L}_\mathrm{clsM} + \lambda \mathcal{L}_\mathrm{PredANN}.
\end{equation}

Here is another intuitive explanation of our model. It utilizes separate encoders for EEG and music data, allowing each encoder to specialize in extracting features specific to its modality for the classification task. Rather than imposing a strict alignment between modalities, we employ a PredANN loss after branching off from the main network that leads to the classification head, preserving unique characteristics of each modality. Moreover, rather than `set-to-set' alignment demonstrated in the prior study \cite{Avramidis2022}, `point-to-point' alignment ensures that the two modalities are aligned at a feature level without distorting the original data representations used for classification (see the Discussion:Previous Paper Comparison section). This alignment strategy demonstrates significant potential for enhancing EEG classification through the incorporation of music data only during training.

\subsubsection*{Model Training and Evaluation} We conducted our training using 6000 epochs to ensure the model convergence. For data extraction during training, we implemented a stride of 200, meaning that during each iteration, data points were extracted every 200 steps, which allowed for efficient use of computational resources by reducing the amount of data processed at each training epoch.  However, for the evaluation of the model's performance, we deviated from this approach. Instead of using the same stride of 200, we evaluated the model using all data points for evaluation with a stride of 1. This means that for the evaluation phase, we used every single data point without skipping any steps. This evaluation approach, using a stride of 1, allows us to obtain a more precise and detailed measure of the model’s accuracy, as it does not omit any data points.

To evaluate the performance of our models, we conducted McNemar's test on each of the seeds. For each prediction, a correct result was marked as 1 and an incorrect result as 0, producing an array of binary outcomes for each model. McNemar's test was then applied to compare the binary arrays from two models, enabling us to statistically assess whether the differences in model predictions were significant. This approach provides a rigorous statistical framework to evaluate model performance beyond mere accuracy, providing a clearer insight into the internal behavior of the models.

\section*{Conclusion}
In this study, we proposed a method for training auditory EEG recognition models by leveraging the similarity between cortical and ANN representations in response to the same auditory stimuli. The effectiveness of this approach was demonstrated in the music identification task, where performance was enhanced by complementing essential information in EEG recordings through training the recognition model to predict ANN representations.

Our experiments showed significant performance improvements for both 1D CNN and 2D CNN architectures in the music identification task. The model exhibited robust learning and effectively incorporated time delays in brain responses to musical stimuli, aligning with recent findings on temporal delays in auditory neural processing. Moreover, we demonstrated that the model could adapt to longer EEG input sequences, with performance improving as the input length increased.

Additionally, we investigated the effects of individual differences and variations in musical stimuli on identification performance. Notably, even individuals with lower overall accuracy in music identification achieved higher performance with specific musical stimuli, emphasizing that differences in the stimuli were more influential than individual differences.

This study demonstrates improved accuracy in recognition models through a framework based on understanding the relationship between audio stimuli and brain recordings. Consequently, it is expected to contribute to the elucidation of cognitive mechanisms through neural decoding, the advancement of brain-computer interfaces (BCI), and the deepening of insights into the relationship between the human brain and ANN representations.

\section*{Data and Code Availability}
\subsection*{Data Availability}
The datasets analyzed during the current study are publicly available \cite{Losorelli2017} and can be accessed via the following link: \url{https://exhibits.stanford.edu/data/catalog/jn859kj8079}.
To note that in the NMED-T dataset we used, each participant's data does not include all songs. For example, participant 14's data lacks trials for song 4. Additionally, the total number of trials varies between participants. However, the total number of trials for each song is equal, with each song having 560 trials. 

\subsection*{Code Availability}
Upon acceptance, the source code utilized for conducting the experiments will be publicly accessible at \url{https://github.com/JURIUENO11/PredANN}.

\bibliography{PredANN_arxiv_v1}

\section*{Author contributions statement}
 T.A. conceptualized and designed the framework, method, and experiment. K.H., P.L., and Z.Z. also designed the method and experiment. Z.Z., K.H., and P.L. implemented the code and conducted the experiment. Z.Z. and P.L. and N.P. analyzed and discussed the results. Z.Z., T.A., and P.L. wrote the main manuscript and created tables and figures. S.M. and Z.Z. organized the code. N.P., T.A., and Z.Z. reviewed the manuscript. N.P. and H.K. advised the research. N.P. organized the research project.

\end{document}